\documentclass[
12pt,
a4paper,
notitlepage,
oneside,
onecolumn]{article}
\usepackage{graphicx}
\usepackage{color}

\begin{document}
\title{On first detection of solar neutrinos from CNO cycle with Borexino}
\author{L.B. Bezrukov, I.S. Karpikov, A.S. Kurlovich, A.K. Mezhokh,\\
 S.V.Silaeva, V.V. Sinev and V.P. Zavarzina }
\maketitle
 Institute for Nuclear Research of Russian academy of sciences, Moscow

\begin{abstract}
Borexino collaboration reported about first measurement of solar CNO-$\nu$ interaction rate in Borexino detector.  
This result is consistent with Hydridic Earth model prediction about the contribution of $^{40}$K geo-antineutrino interactions in single Borexino events.
The potassium abundance in the Earth in the range $1 \div 1.5$\% of the Earth mass could give the observed enhancement of counting rate above expected CNO-$\nu$ counting rate. The Earth intrinsic heat flux must be in the range $200 \div 300$ TW for this potassium abundance. This value of the heat flux can explain the ocean heating observed by the project ARGO. 
We consider that Hydridic Earth model actually corresponds better to CNO-$\nu$ Borexino results than Silicate Earth model.
\end{abstract} 

\section{Introduction.}
\hspace{0.5cm}

Recently the Borexino collaboration \cite{Ranucci2020} reported at Neutrino2020 the results of first detection of solar neutrinos from CNO cycle and published the arXiv \cite{agost2020}. They used of two types of analysis of experimental data: Counting Analysis (CA) and Multivariate fit (MF).

We will use the following notations below. 

Differential energy spectrum of recoil electron counting rate:
\begin{equation}
R'(E) = \frac{d R(E)}{dE}.
\end{equation}
Here and below E - the energy of recoil electron in the scattering of neutrino (or antineutriono) on electron.

The recoil electron counting rate in the energy range $E_{min} \div E_{max}$ :
\begin{equation}
 R(E_{min} \div E_{max}) = \int_{E_{min}}^{E_{max}} R'(E) dE.
\end{equation} 

The integral interaction rate without energy threshold:
\begin{equation}
 R = \int_0^\infty R'(E) dE.
\end{equation}

We will express the $R(E_{min} \div E_{max})$ and $R$ in units: cpd/100 t - counts per day in 100 tonnes of scintillator.

Expected integral interaction rate from Standard Solar Model under the high metallicity (SSM HM) and MSW-LMA effect for CNO-$\nu$  is \cite{agost2020}:
\begin{equation}\label{eq:3}
R_{\nu,theory} = 4.92 \pm 0.78 \ \rm{cpd/100\ t}. \hspace{0.7cm} (68\% C.L.)
\end{equation}

The Counting Analysis used the energy range from $E_{min} = 0.74$ MeV to $E_{max} = 0.85$ MeV. The integral interaction rate (without energy threshold) of CNO-$\nu$ was obtained by the following way: 
\begin{equation}\label{eq:1}
R_{\nu,CA} = \frac{R_{\nu,CA}(0.74\div0.85 \ \rm{MeV})}{R_{\nu,theory}(0.74\div0.85 \ \rm{MeV})}\cdot R_{\nu,theory} = 5.6 \pm 1.6 \ \rm{cpd/100 \ t} \hspace{0.4cm} (68\% C.L.),
\end{equation}
where $R_{\nu,CA}(0.74\div0.85\rm{MeV})$ is experimentally attributed as CNO-$\nu$ event counting rate in the energy range  $0.74 \div 0.85$ MeV.

Multivariate fit utilizes the energy spectrum shapes of CNO-$\nu$ signal and its backgrounds. The Multivariate fit used the energy range from $E_{min} = 0.32$ MeV to $E_{max} = 2.64$ MeV.  The best fit gives the result: 
\begin{equation}\label{eq:2}
R_{\nu,MF} = \frac{R_{\nu,MF}(0.32\div2.64\ \rm{MeV})}{R_{\nu,theory}(0.32\div2.64\ \rm{MeV})}\cdot R_{\nu,theory} = 7.2 - 1.7 + 3.0 \ \rm{cpd/100\ t} \hspace{0.3cm} (68\% C.L.),
\end{equation}
where $R_{\nu,MF}(0.32\div2.64\ \rm{MeV})$ is experimentally attributed as CNO-$\nu$ event counting rate in the energy range  $0.32 \div 2.64$ MeV as a result of the best fit.

The Borexino claims \cite{Ranucci2020} the detection of CNO-$\nu$ flux with significance 5$\sigma$. This is the remarkable result.

We predicted \cite{BezrarXiv20} by using Hydridic Earth model \cite{Larin, toul} that the total interaction rate of the solar CNO-$\nu$ and $^{40}$K geo-antineutrino will be obtained in the range of $6 \div 9$ cpd/100 t.

\vspace{2cm}
\section{Hypothesis of new source of single Borexino events}
\hspace{0.5cm}

We will pay attention here on the relations of the most probable values for different types of analysis of experimental data: Counting Analysis (CA) and Multivariate fit (MF): 
\begin{equation}\label{eq:4}
\frac{R_{\nu,CA}}{R_{\nu,theory}} = \frac{R_{\nu,CA}(0.74\div0.85 \ \rm{MeV})}{R_{\nu,theory}(0.74\div0.85 \ \rm{MeV})} = \frac{5.6}{4.92} = 1.138. 
\end{equation}
\begin{equation}\label{eq:5}
\frac{R_{\nu,MF}}{R_{\nu,theory}} = \frac{R_{\nu,MF}(0.32\div2.64\ \rm{MeV})}{R_{\nu,theory}(0.32\div2.64\ \rm{MeV})} = \frac{7.2}{4.92} = 1.463. 
\end{equation}

We can see that we need the new source of single events with soft energy specrtum compearing CNO-$\nu$ energy spectum to explain numbers (\ref{eq:4}) and (\ref{eq:5}).  

The authors of the works \cite{bezr, BezrarXiv20} considered an additional source of single events for the Borexino detector. This is the scattering of potassium geo-antineutrinos on electrons.  

 Such a source was not considered in \cite{Ranucci2020, agost} because the Silicate Earth model was used in it. The abundance of potassium in this model is very small (0.024\% of the Earth mass) and the contribution to single events of the Borexino detector from potassium geo-antineutrinos is negligible.

But in the works \cite{bezr} and \cite{BezrarXiv20} the authers used Hydridic Earth model which predicted the abundance of potassium in the Earth up to 4\% of the Earth mass.

We present in Table 1 the differential energy spectrum of counting rate of recoil electrons $R'_{\nu,theory}(E)$ from CNO-$\nu$ scattering in 100 tonnes of scintillator (column 2). The total differential energy spectra of counting rate of recoil electrons from the scattering of CNO-$\nu$ and $^{40}$K geo-antineutrino for the abundances of potassium in the modern Earth  1\% and 1.5\% of the Earth mass  are in (column 3) and  (column 4) correspondingly. The neutrino (and antineutrino) oscillations were taken into account.

The intensity of events from the CNO-$\nu$ calculated for the function from column 2 is $R_{\nu,theory}$ = 4.9 cpd/100 t.

 The intensity of events for the function from column 3 with $^{40}$K geo-antineutrinos contribution is 

$R_{\rm{CNO}+1\%\rm{K}}$ = 7.05 cpd/100 t.

The intensity of events for the function from column 4 with $^{40}$K geo-antineutrinos contribution is 

$R_{\rm{CNO}+1.5\%\rm{K}}$ = 8.1 cpd/100 t.

\begin{table}[ht]
\caption{Differential energy spectra of counting rate of recoil electrons from solar CNO-$\nu$ scattering on electrons in 100 tonnes of scintillator and the same with addition of $^{40}$K geo-antineutrinos in units MeV$^{-1}$ year$^{-1}$ (100\ tonnes)$^{-1}$. The neutrino (and antineutrino) oscillations were taken into account.}
\label{tabl:event}
\centering
\vspace{2mm}
 \begin{tabular}{| c | c | c | c |c |} 
 \hline
 E, MeV &  CNO  & CNO+1\%K  & CNO+1.5\%K & CNO+2\%K \\
 \hline
 0.1 & 2997.6 & 5254.4 & 6334.6 & 7414.8 \\
 0.2 & 2736.4 & 4363.7 & 5142.6 & 5921.6 \\
 0.3 & 2447.1 & 3597.8 & 4148.6 & 4699.4 \\
 0.4 &2129.8 & 2924.4 &3304.8 & 3685.1 \\
 0.5 & 1787.8 & 2319.7 & 2574.2 & 2828.8 \\
 0.6 & 1439.3 & 1781.1 & 1944.8 & 2108.4 \\
 0.7 & 1100.8 & 1308.2 & 1407.4 & 1506.6 \\
 0.8 & 787.7 & 901.9 & 956.5 & 1011.2 \\
 0.9 & 526.8 & 579.6 & 604.8 & 630.0 \\
 1.0 & 355.7 & 371.5 & 379.0 & 386.5 \\
 1.1 & 249.9 & 249.9 & 249.9 & 249.9 \\
 1.2 & 156.7 & 156.7 & 156.7 & 156.7 \\
 1.3 & 81.17 & 81.17 & 81.17 & 81.17 \\
 1.4 & 26.75 & 26.75 & 26.75 & 26.75 \\
 1.5 & 0.303 & 0.303 & 0.303 & 0.303 \\
 \hline
\end{tabular}
\vspace{7 mm}
\end{table}

 Let calculate the ratios from Table 1 similar to (\ref{eq:4}) and (\ref{eq:5}) for the same energy ranges:
\begin{equation}\label{eq:6}
\frac{R_{\nu,\rm{CNO}+1\%\rm{K}}(0.74\div0.85 \ \rm{MeV})}{R_{\nu,theory}(0.74\div0.85 \ \rm{MeV})}
 \div  \frac{R_{\nu,\rm{CNO}+1.5\%\rm{K}}(0.74\div0.85 \ \rm{MeV})}{R_{\nu,theory}(0.74\div0.85 \ \rm{MeV})} = 1.16 \div 1.25.
\end{equation}
\begin{equation}\label{eq:7}
\frac{R_{\nu,\rm{CNO}+1\%\rm{K}}(0.32\div2.64 \ \rm{MeV})}{R_{\nu,theory}(0.32\div2.64 \ \rm{MeV})}
 \div  \frac{R_{\nu,\rm{CNO}+1.5\%\rm{K}}(0.32\div2.64 \ \rm{MeV})}{R_{\nu,theory}(0.32\div2.64 \ \rm{MeV})} = 1.29  \div 1.43.   
\end{equation}

Let compare (\ref{eq:4}) with (\ref{eq:6}) and (\ref{eq:5}) with (\ref{eq:7}). We can see that the values are close. 

We can conclude that the potassium abundance in the Earth in the range $1 \div 1.5$\% of the Earth mass could explain the observed enhancement of counting rate above expected CNO-$\nu$ counting rate. We can see that Hydridic Earth model helps to explain the Borexino results of detection of CNO-$\nu$ flux.

\vspace{2cm}
\section{Chi test of Hypotheses.}
\hspace{0.5cm}

We will calculate the following values to compare the validity of  different models of CNO like events:

\vspace{0.5 cm}
${\large\chi_i = \chi_{MF,i} + \chi_{CA,i} = }$  
\begin{eqnarray}
\frac{|R_{\nu,MF}(0.32\div2.64\ {\rm MeV}) - R_{model,i}(0.32\div2.64\ {\rm MeV})|}{\sigma_{MF}}\nonumber\\
+ \hspace{0.3cm}\frac{|R_{CA}(0.74\div0.85\ {\rm MeV}) - R_{model,i}(\ {0.74\div0.85\rm MeV})|}{\sigma_{CA}},
\end{eqnarray}
where $i$ - the index of different models (see Table 2).

\begin{table}[ht]
\caption{Different models of CNO like events. Index of model is shown in column 1.} 
\label{tabl:models}
\centering
\vspace{2mm}
 \begin{tabular}{| c | c | c |} 
 \hline
 $i$ &  Model  & $\chi_{MF,i} + \chi_{CA,i} =\chi_i$ \\ 
 \hline
 1 & CNO energy spectrum, $R$ = 4.9 cpd/100 t&1.33 + 0.44 = 1.77 \\
 2 & CNO energy spectrum, $R$ = 5.6 cpd/100 t&0.94 + 0.0 = 0.94 \\
3 & CNO energy spectrum, $R$ = 7.2 cpd/100 t&0.0 + 1.0 = 1.0. \\
 4 & CNO+1\%K energy spectrum,  $R$ = 7.05 cpd/100 t&0.51 + 0.0 = 0.51 \\
 5 &CNO+1.5\%K energy spectrum, $R$ = 8.1 cpd/100 t&0.11 + 0.22 = 0.33\\
6 &CNO+2\%K energy spectrum, $R$ = 9.2 cpd/100 t&0.13 + 0.43 = 0.56 \\
\hline
\end{tabular}
%\vspace{10mm}
\end{table}

If $R_{\nu,MF}(0.32\div2.64\ {\rm MeV}) - R_{model,i}(0.32\div2.64\ {\rm MeV}) > 0$ we can wright using (6):

$\sigma_{MF} = R_{\nu,MF}(0.32\div2.64\ {\rm MeV}) \cdot 1.7 / 7.2$.

If $R_{\nu,MF}(0.32\div2.64\ {\rm MeV}) - R_{model,i}(0.32\div2.64\ {\rm MeV}) < 0$ we can wright using (6):

$\sigma_{MF} = R_{\nu,MF}(0.32\div2.64\ {\rm MeV}) \cdot 3.0 / 7.2$.

We calculated $\sigma_{CA}$ by the same method using (5):

  $\sigma_{CA} = R_{CA}(0.74\div0.85\ {\rm MeV}) \cdot 1.6 / 5.6$.

We used the values from Table 1 for calculation of {\large$\chi_i$}. The results are shown in Table 2 (column 3).  The closer {\large$\chi$} to zero the model is better.
We see that model 5 is preferable. This is CNO-$\nu$  from SSM HM and MSW-LMA effect plus $^{40}$K geo-antineutrino for 1.5\% abundance of K in the Earth with resulting $R$ = 8.1 cpd/100 t. Note that such abundance of K is predicted by Hydridic Earth model throughout the Earth. The Silicate Earth model predicts such abundance of K only in the Earth crust and the absence of K in the Earth mantle.

\section{Terrestrial heat production.}

\hspace{0.5cm}

Let calculate the intrinsic Earth heat flux in the frame of Hydridic Earth model for the potassium abundance 1\% of the Earth mass.
The mass of $^{40}$K in the Earth is:
$m(^{40}\rm{K}) = 0.7 \cdot 10^{22}$ g.

The equation relating masses and heat production is
\begin{equation}\label{eq:8}
H = m \cdot \frac{N_{Avog}}{A} \cdot \frac{E_{release}}{\tau} \cdot \alpha, 
\end{equation}
	where $N_{Avog}$ - Avogadro number, $A$ - atomic number, $E_{release}=0.6$ MeV - average energy release in $^{40}$K decay, $\tau = t_{1/2} / ln2$ - mean lifetime of isotope, $\alpha$ - the conversion factor $1\ \rm{MeV} = 1.6 \cdot 10^{-13}$ J
\begin{equation}\label{eq:9}
H(^{40}\rm{K}) = 0.7\cdot 10^{22}\rm{ g} \frac{6 \cdot 10^{23}}{40} \rm{g^{-1}} \cdot \frac{0.6\ MeV} {1.8 \cdot 10^9 \cdot 3.15 \cdot 10^7 s} \cdot 1.6 \cdot 10^{-13} J = 177 \ TW. 
\end{equation}
The Hydridic Earth model predicts the abundance of U and Th some larger comparing with prediction of Silicate Earth model.  The estimation of total intrinsic earth heat flux is:
\begin{equation}\label{eq:10}
H = H({\rm{U}}) + H({\rm{Th}}) + H(^{40}\rm{K}) = 200 \ TW. 
\end{equation}

The similar estimation for the potassium abundance 1.5\% of the Earth mass gives the value about 300 TW.

The obtained range of heat flux $H \sim 200 \div 300$ TW can explain the observed by ARGO project the heating of the ocean \cite{Argo}.

\section{Reasons to use the Hydridic Earth model.}

\vspace{0.5cm}
\hspace{0.5cm} The widespread belief in the fairness of Silicate Earth model and belief in the validity of the results of work \cite{davis} that the heat flux from the Earth interior is equal to 47$\pm$2 TW do not allow to explain the CNO solar neutrinos interaction rate obtained by Borexino.

Often the Hydridic Earth model is critisized by the following points. The entire Earth could have melt due to radiogenic heat and spent the most part of its life in this state if the Earth contains potassium more than 1\% of the Earth weight. Also the Hydridic Earth model predicts that the current heat  flux from the Earth's interior can be 200 TW and more which contradicts to the result of the work \cite{davis}.  

However,
these arguments are not fully reliable.  In particular, the entire Earth could not have melt because the Hydridic Earth model contains a subsurface cooling mechanism. This mechanism is activated when the subsurface is heated enough to decompose the metal hydrides. Therefore, in the Hydridic Earth model, the subsurface temperature oscillates \cite{Larin} and not grows up till hydrides exist in the Earth. This argument was used in \cite{bezr2, bezr2018, barab2019, bezr22018}. It is  noted in these works  that thermal conductivity is not the main mechanism of heat transfer in the Earth, but protons and hydrogen-containing gases carry out the heat away. In these works the experimental evidences are provided that the heat flux from Earth interior can reach the several hundreds TW. These are the heating of the oceans  \cite{Argo}, the temperature profile of ultra-deep wells and non-direct evidence – Moon heat flux from interior.

Moreover, the effect was found that is predicted by the Hydridic Earth model and not predicted by the Silicate Earth model.
The Hydridic Earth model predicts that the Earth's crust is positively charged and contains a large amount of positive charge in the form of protons and positive ions of various hydrogen-containing gases. We tested the validity of this prediction experimentally by detecting an excess of the concentration of positive air-ions in underground rooms over the concentration of negative air-ions. In order to make this prediction, a new model of terrestrial electricity was developed based on the Hydridic Earth model \cite{bezr2019}. The model named "Hydridic model of terrestrial electricity".  This model explains the origin of the atmospheric electric field and all the observed effects of atmospheric electricity in a single way. The model also explains the origin of telluric currents. In \cite{bezr22019}, the Hydridic model of terrestrial electricity was successfully used in the analysis of the reaction of telluric currents to earthquakes.

We consider that Hydridic Earth model actually corresponds better to CNO-$\nu$ Borexino results than Silicate Earth model. It is CNO-$\nu$  from SSM HM and MSW-LMA effect plus $^{40}$K geo-antineutrino for 1.5\% abundance of K in the Earth with resulting $R$ = 8.1 cpd/100 t.

\vspace{2cm}
 
\section{Conclusion.}
\begin{enumerate}
\item 
We propose the most probable interpretation of CNO Borexino results reported on Neutrino2020 as the existence of new source of single events with more soft energy spectrum comparing with energy spectrum of recoil electrons from CNO-$\nu$ interactions.
\item 
We propose to consider the scattering of $^{40}$K geo-antineutrinos on electrons as new source of single Borexino events predicted by Hydridic Earth model. 
\item
We propose to include the scattering of $^{40}$K geo-antineutrinos by electrons in the analysis of single event energy spectrum of Borexino detector with the intensity of $^{40}$K geo-antineutrinos flux as a free parameter. 
\item
The Chi test of models of different origins of CNO like events allow to choose the most probable model. It is CNO-$\nu$  from SSM HM and MSW-LMA effect plus $^{40}$K geo-antineutrino for 1.5\% abundance of K in the Earth with resulting $R$ = 8.1 cpd/100 t.
\end{enumerate}

\section{Acknowledgment}
The authors express their gratitude to F.L.Bezrukov and I.I.Tkachev for the interest and the helpful remarkes.


\begin{thebibliography}{99}

\bibitem{Ranucci2020} G.Ranucci {\it et al.} (Borexino collaboration), First detection of solar neutrinos from CNO sycle with Borexino, report on Neutrino2020 conference.
https://indico.fnal.gov/event/43209/contributions/187871/attachments
/129210/158592/borexino\_cno\_neutrino2020.pdf

\bibitem{agost2020} M. Agostini and others,
    collaboration BOREXINO, First Direct Experimental Evidence of CNO neutrinos, arXiv: 2006.15115 [hep-ex]

\bibitem{BezrarXiv20} L.B.~Bezrukov, I.S.~Karpikov, A.S.~Kurlovich, A.K.~Mezhokh, S.V.~Silaeva, V.V.~Sinev, V.P.~Zavarzina, On the contribution of the $^{40}$K geo-antineutrino to single Borexino events, 
 arXiv:2004.02533v2 [hep-ex] 

\bibitem{agost}  M.~Agostini, K.~Altenmuller, S.~Appel {\it et al.}, Simultaneous precision spectroscopy of {\it pp}, $^{7}$Be, and {\it pep} solar neutrinos with Borexino Phase-II, Phys. Rev. \textbf{D 100} (2019) 082004, doi:10.1103/PhysRevD.100.082004; arXiv:1707.09279 [hep-ex].

\bibitem{bezr} V.V.~Sinev, L.B.~Bezrukov, E.A.~Litvinovich, I.N.~Machulin, M.D.~Skorokhvatov, Looking for Antineutrino Flux from $^{40}$K with Large Liquid Scintillator Detector, Physics of Particles and Nuclei \textbf{46} (2015) 186, doi:10.1134/S1063779615020173; arXiv:1405.3140 [physics.ins-det].

\bibitem{Larin}
V.N. Larin, Hydridic Earth: the New Geology of Our Primordially Hydrogen-Rich Planet. / Ed. C. Warren Hunt. Calgary, Alberta, Canada: Polar Publishing, 1993. 247p.

\bibitem{toul} Herve Toulhoat, Valerie Beaumont, Viacheslav Zgonnik, Nikolay Larin, Vladimir N. Larin, Chemical differentiation of planets: a core issue, arXiv:1208.2909 [astro-ph.EP]. 

\bibitem{Argo} Riser, S.C., H.J. Freeland, D. Roemmich, et al, Fifteen years of ocean observations with the global Argo array. Nature Clim. Change, 2016, Vol. 6, 145-153, http://dx.doi.org/10.1038/nclimate2872

\bibitem{davis} J.H.~Davies and D.R.~Davies, Earth’s surface heat flux, Solid Earth \textbf{1} (2010) 5, doi:10.5194/se-1-5-2010.

%\bibitem{Bahcall}J. N. Bahcall and R. K. Ulrich, Rev. Mod. Phys. 60, 297 (1988). J. N. Bahcall, https://www.sns.ias.edu/$\sim$ jnb/SNdata/cnospectra.html (2005).

\bibitem{bezr2} L.B.~Bezrukov, A.S.~Kurlovich, B.K.~Lubsandorzhiev, A.K.~Mezhokh, V.P.~Morgalyuk, V.V.~Sinev and V.P.~Zavarzina, How Geoneutrinos can help in understanding of the Earth heat flux, J. Phys. Conf. Ser. \textbf{934} no.1 (2017) 012011, doi:10.1088/1742-6596/934/1/012011.

\bibitem{bezr2018} L. B. Bezrukov, A. S. Kurlovich, B. K. Lubsandorzhiev, A. K. Mezhokh, V. P. Morgalyuk, V. V. Sinev, and V. P. Zavarzina, Geo-Neutrinos and the Earth’s Internal Heat Flux. Physics of Particles and Nuclei, 2018, Vol. 49, No. 4, pp. 674–677. Original Russian Text © L.B. Bezrukov, A.S. Kurlovich, B.K. Lubsandorzhiev, A.K. Mezhokh, V.P. Morgalyuk, V.V. Sinev, V.P. Zavarzina, 2018, published in Fizika Elementarnykh Chastits i Atomnogo Yadra, 2018, Vol. 49, No. 4, pp.
 https://doi.org/10.1134/S1063779618040135
\bibitem{barab2019}
I.R. Barabanov, L.B. Bezrukov, V.P. Zavarzina, I.S. Karpikov, A.S. Kurlovich, B.K. Lubsandorzhiev, A.K. Mezhokh, V.P. Morgalyuk, V.V. Sinev, Study of Earth’s Heat Flux by Means of Geoneutrino Detection. 
Published in Phys.Atom.Nucl. 82 (2019) no.1, 8-12 \\
DOI: 10.1134/S1063778819010034
\bibitem{bezr22018}L. B. Bezrukov, A. S. Kurlovich, B. K. Lubsandorzhiev, V. V. Sinev, V. P. Zavarzina and V. P. Morgalyuk. Geo-neutrino, Earth heat flux, Earth electricity. 
 EPJ Web of Conferences 191, 03005 (2018) 	
QUARKS-2018. \\
DOI: https://doi.org/10.1051/epjconf/201819103005

\bibitem{bezr2019}L.B. Bezrukov, V.P. Zavarzina, A.S. Kurlovich, B.K. Lubsandorzhiev, A.K. Mezhokh, V.P. Morgaluk, V.V. Sinev, On Negatively Charged Layer of the Earth’s electric Field //Doklady Physics, 2018, Vol. 63, No. 5, pp. 177–179. Original Russian Text © L.B. Bezrukov, V.P. Zavarzina, A.S. Kurlovich, B.K. Lubsandorzhiev, A.K. Mezhokh, V.P. Morgaluk, V.V. Sinev, Doklady Akademii Nauk, 2018, Vol. 480, No. 2, pp. 155–157. https://doi.org/10.1134/S1028335818050051
\bibitem{bezr22019}L. B. Bezrukov, I. S. Karpikov, A. S. Kurlovich, B. K. Lubsandorzhiev, A. K. Mezokh, V. P. Morgaluk, V.V. Sinev and V. P. Zavarzina. INTERPRETATION OF RESULTS OF MEASUREMENTS OF VOLTAGE INTO BAYKAL LAKE. Geomagnetism and Aeronomy. 2019. V.59. No 5. pp 623-627. 
DOI: 10.1134/S0016793219040054
Original Russian Text © L.B. Bezrukov, V.P. Zavarzina, I.S.Karpikov, A.S. Kurlovich, B.K. Lubsandorzhiev, A.K. Mezhokh, V.P. Morgaluk, V.V. Sinev, Geomagnetizm I Aeromomiya, 2019, Vol. 59, No. 5, pp. 666-670. \\
DOI: 10.1134/S0016794019040059

\end{thebibliography}
\end{document}